\documentclass[letter, twoside, journal]{IEEEtran}

\usepackage{cite}
\usepackage[pdftex]{graphicx}
\usepackage{amsmath}
\usepackage[caption=false,font=footnotesize]{subfig}

\usepackage[dvipsnames]{xcolor}

\usepackage{multirow}
\usepackage{float}

\usepackage{tikz}

\usetikzlibrary{decorations.pathreplacing,calligraphy}

\usepackage[utf8]{inputenc}
\usepackage{pgfplots} 
\usepackage{pgfgantt}
\usepackage{pdflscape}
\pgfplotsset{compat=newest} 
\pgfplotsset{plot coordinates/math parser=false}

\usepackage{siunitx}
\DeclareSIUnit{\nothing}{\relax}

\usepackage{newtxtext,newtxmath}

\usetikzlibrary{shapes}

\usepackage{bm}

\usepackage{booktabs}

\usepackage{blindtext}

\usepackage{upgreek}

\usepackage{accents}

\usepackage{array}
\newcolumntype{L}[1]{>{\raggedright\let\newline\\\arraybackslash\hspace{0pt}}m{#1}}
\newcolumntype{C}[1]{>{\centering\let\newline\\\arraybackslash\hspace{0pt}}m{#1}}
\newcolumntype{R}[1]{>{\raggedleft\let\newline\\\arraybackslash\hspace{0pt}}m{#1}}

\newlength\fwidth
\newlength\fheight

\begin{document}
\title{Characterization and Modeling of Self-Heating in Nanometer Bulk-CMOS at Cryogenic Temperatures}

\author{\IEEEauthorblockN{P.~A.~'t Hart$^{1,2}$, M.~Babaie$^{1}$, A.~Vladimirescu$^{1,2,3,4}$ and F.~Sebastiano$^{1,2}$}

\IEEEauthorblockA{$^1$QuTech, Delft University of Technology, The Netherlands; $^2$Department of Quantum and Computer Engineering, Delft University of Technology, The Netherlands; $^3$ISEP, Paris, France; $^4$UC Berkeley, Berkeley, CA, USA}
}

\maketitle

\begin{abstract}
This work presents a self-heating study of a 40-nm bulk-CMOS technology in the ambient temperature range from \SI{300}{\kelvin} down to \SI{4.2}{\kelvin}. A custom test chip was designed and fabricated for measuring both the temperature rise in the MOSFET channel and in the surrounding silicon substrate, using the gate resistance and silicon diodes as sensors, respectively. Since self-heating depends on factors such as device geometry and power density, the test structure characterized in this work was specifically designed to resemble actual devices used in cryogenic qubit control ICs. Severe self-heating was observed at deep-cryogenic ambient temperatures, resulting in a channel temperature rise exceeding \SI{50}{\kelvin} and having an impact detectable at a distance of up to \SI{30}{\micro \meter} from the device. By extracting the thermal resistance from measured data at different temperatures, it was shown that a simple model is able to accurately predict channel temperatures over the full ambient temperature range from deep-cryogenic to room temperature. The results and modeling presented in this work contribute towards the full self-heating-aware IC design-flow required for the reliable design and operation of cryo-CMOS circuits.
\end{abstract}

\section{Introduction}
\begin{figure}[t!]
\setlength\fwidth{0.35\textwidth}
\centering
\hspace{0mm}
\includegraphics[angle=0]{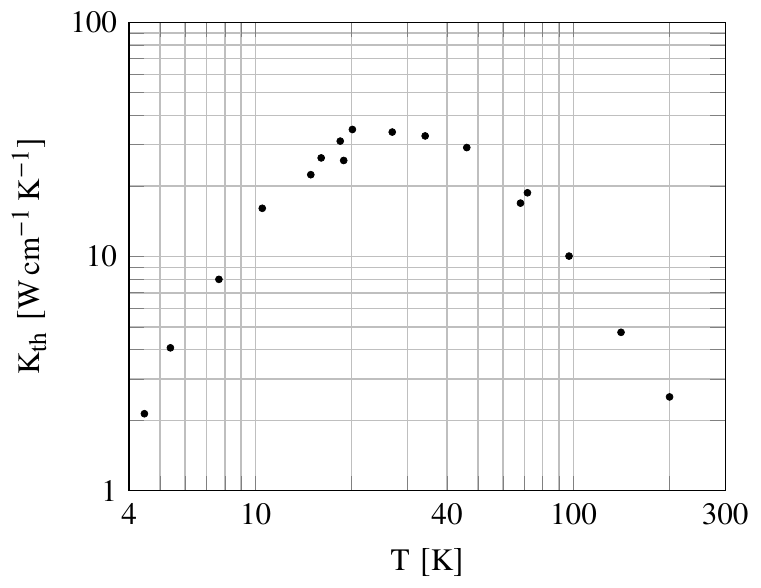}
\vspace{-1mm}
\caption{Thermal conductivity of silicon over temperature, replotted from~\cite{Glassbrenner64}.}
\label{fig_CondSi}
\vspace{-4.5mm}
\end{figure}

Quantum computers have the potential to solve certain computational problems that would otherwise take a prohibitive long time to complete using classical computers. For proper operation, the quantum bits (qubits) --the basic unit of information in quantum computers-- need to be cooled down to deep-cryogenic temperatures, around a few Kelvin in some cases~\cite{Vandersypen17} but typically below \SI{100}{\milli \kelvin}~\cite{Vandijk19}. Since state-of-the-art quantum computers comprise only a handful of qubits, each qubit can be individually wired to equipment placed at room temperature (RT)\cite{Google19}. However, to be of any practical use, future quantum computers require thousands to even millions of physical qubits, making today's  approach unworkable due to the need for thousands of cables going from the cryogenic qubits to the RT equipment. The problems associated with scalability, manufacturability and reliability of these systems could be solved by placing integrated control electronics in close vicinity to the qubits, thus requiring electronic circuits operating at cryogenic temperatures. Such electronics are typically operated at liquid helium (LHe, \SI{4.2}{\kelvin}) temperature, as commonly-adopted dilution refrigerators can offer significant cooling power ($\approx$ $\SI{1}{\watt}$)~\cite{Sebastiano17} only at those temperatures.\\
The technology of choice for the cryogenic controller is nanometer CMOS, for its high speed, maturity, integration density and its potential to operate down to \SI{30}{\milli \kelvin}~\cite{Incandela18,Ekanayake10}, all required for handling a large number of qubits.\\
It has been shown that core device parameters, such as threshold voltage, mobility, subthreshold slope, mismatch and leakage, can shift significantly from their RT values at low ambient temperatures ($T_{amb}$)~\cite{Beckers18,Hart20_1,Hart20_2,Incandela18,Patra20}. The incorporation of device temperature in compact models extended to the cryogenic environment is thus paramount to guarantee reliable circuit simulations and hence, robust circuit operation under these conditions. However, self-heating (SH) can raise the device temperature ($T_{chan}$) significantly above $T_{amb}$. This effect is amplified at cryogenic temperatures, as thermal properties of silicon, such as thermal conductivity ($K_{th}$), vary almost over 1.5 orders of magnitude in the temperature range from RT down to \SI{4.2}{\kelvin}, as shown in Fig.~\ref{fig_CondSi}.
For instance, a recent cryo-CMOS microwave driver for spin qubits operating at $T_{amb}$ = \SI{3}{\kelvin} was subjected to SH exceeding \SI{10}{\kelvin} for a dissipated power above \SI{400}{\milli \watt}~\cite{Vandijk20}.\\
SH does not only impact the characteristics of the device itself, it can also propagate through the surrounding silicon forming thermal feedback loops with neighboring devices~\cite{Gutierrez93}. While this can be already critical in electronic cryogenic circuits, it will become crucial in future system-on-chip (SoC), integrating both electronics and qubits, which are extremely sensitive to any thermal crosstalk~\cite{Vandersypen17}.\\
SH at RT has received much attention in literature, specifically focused on silicon-on-insulator (SOI) technologies as the buried oxide (BOX) poses a thermal impedance 2 orders of magnitude higher compared to that of bulk silicon at this temperature~\cite{Glassbrenner64,Asheghi98}.\\
Far less attention was devoted to studies on SH at cryogenic temperatures. Early work dates back to the beginning 1970s \cite{Sesnic72} and has been extended more recently by investigations on bulk MOSFETs~\cite{Sesnic72,Foty87,Foty89,Gutierrez93,Artanov21}, resistors~\cite{Gutierrez93_2,Hidalga00} and SOI~\cite{Jomaah95,Triantopoulos19}. SH was investigated both by measurements of the device temperature itself~\cite{Jomaah95,Triantopoulos19,Hidalga00} and by placing temperature sensors in the vicinity of on-chip heaters~\cite{Gutierrez93,Gutierrez93_2,Hidalga97,Gutierrez92,Gutierrez97,Hidalga00}. All these works show that SH is exacerbated at cryogenic temperatures and that the effect is highly dependent on device geometry (size, aspect ratio) and power density. This variability is clearly observed in recent cryo-CMOS integrated circuits for qubit interfacing, as SH ranged from 1 to \SI{3}{\kelvin} in a 40-nm bulk-CMOS high-speed ADC~\cite{Kiene21} to more than \SI{10}{\kelvin} in a 22-nm FinFET microwave driver~\cite{Vandijk20}. As device geometry and power density differ considerably between advanced bulk CMOS nodes and the previously studied mature technologies, it is necessary from a modeling perspective to characterize SH on devices  better resembling those employed in practical cryo-CMOS designs~\cite{Prabowo21,Kiene21,Ruffino21,Park21}, both in geometry and power density. Understanding the impact of SH is especially  crucial for the cryo-CMOS low-noise amplifiers (LNA) necessary for the detection of the weak signals from quantum processors, as an increase of the device temperature of only a few Kelvin can strongly affect the noise performance, e.g., in a thermal-noise-limited amplifier in which the noise is directly proportional to the device temperature.\\
This paper bridges this gap by characterizing and modeling the effects of SH on the device itself and on the surrounding silicon, using a typical NMOS device.\\
It is found that SH can have a severe impact on both the device operating temperature and the temperature of the surrounding silicon at deep-cryogenic temperatures, and, that the former effect can be be successfully predicted using a simple modeling approach.\\
This paper is structured as follows: Section~\ref{sect_TestStructuresAndMeasurementSetup} describes the test chip, measurement setup and the device calibration. Section~\ref{sect_MeasurementsResultsAndAnalysis} presents the measurement results, which are discussed and modeled in Section~\ref{sect_Discussion}. Finally, conclusions are drawn in Section~\ref{sect_Conclusion}.

\section{Test Structures and Measurement Setup}
\label{sect_TestStructuresAndMeasurementSetup}
\begin{figure}[t!]
\setlength\fwidth{0.20\textwidth}
\centering
\hspace{1mm}
\includegraphics[angle=0]{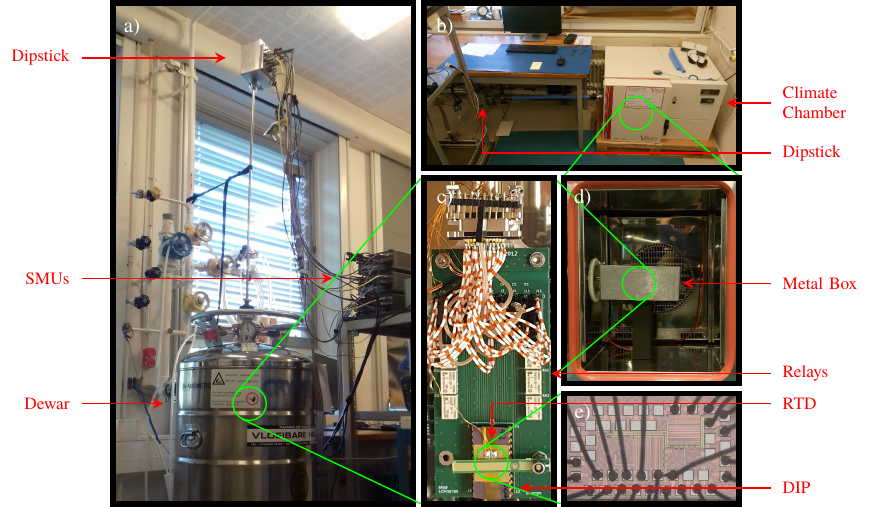}
\vspace{-2mm}
\caption{Measurement setup: a) dipstick in LHe dewar; b) dipstick in climate chamber; c) PCB at the end of the dipstick; d) climate chamber internal view; e) die micrograph.}
\label{fig_Setup}
%\vspace{-4.5mm}
\end{figure}

A test chip was taped-out, specifically designed for the characterization of SH at deep-cryogenic temperatures. The chip was manufactured in the TSMC 40-nm bulk-CMOS process. Fig.~\ref{fig_Structures} and Fig.~\ref{fig_Setup}e show a simplified overview of the test structures and a die micrograph, respectively.\\
Three NMOS devices are employed as heaters (H1, H2 and H3), formed by a 5 fingered device with fingers measuring $W/L=\SI{12}{\micro \meter}/\SI{40}{\nano \meter}$ each, individually selectable (separated gates and drains) and able to dissipate $\approx \SI{7}{\milli \watt}$ of power. The gates of the two MOSFETs separating H1 and H3 from H2 are connected to $\mathrm{V_{SS}}$ in order to electrically isolate the heaters from each other (Fig.~\ref{fig_Structures} top). The center heater (H2) has additional connections available, enabling the measurement of the gate resistance, further discussed in Section~\ref{sect_GateTestStructure}. The choice for NMOS over a PMOS device was motivated by its higher current driving capability (and thus power), as no significant thermal differences are expected between both types.\\
In addition to the MOSFETs, a linear array comprising 52 diodes is placed perpendicular to the channel, along a line through the center of the heaters (Fig.~\ref{fig_Structures} top). These diodes act as temperature sensors, enabling the detection of the spatial thermal profile in the heaters' vicinity, further discussed in Section~\ref{sect_DiodeTestStructures}.
\subsection{Gate Test Structure}
\label{sect_GateTestStructure}%
\begin{figure}[t!]
\setlength\fwidth{0.30\textwidth}
%\centering
\hspace{1mm}
\includegraphics[angle=0]{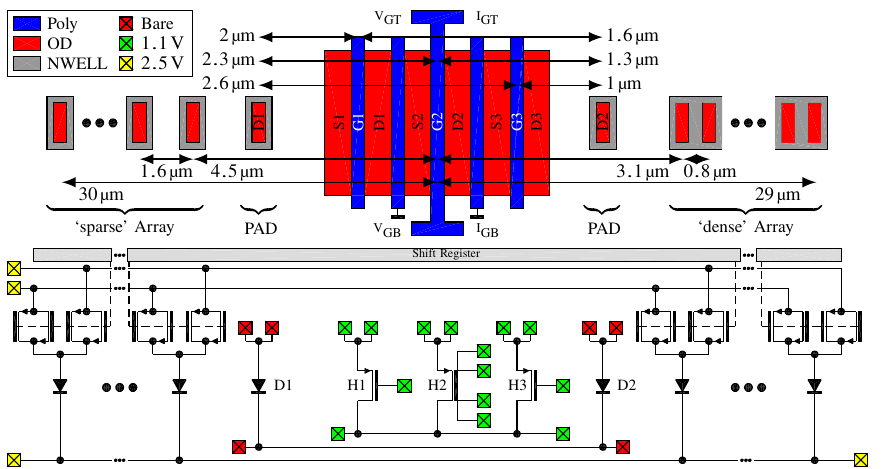}
\vspace{-3mm}
\caption{Simplified layout (top) and schematic overview (bottom) of the gate and diode test structures. H2 has Kelvin connections to its gate, comprising the $V_{GT}$, $I_{GT}$, $V_{GB}$ and $I_{GB}$ contacts. D1 and D2 are the two pad-accessible diodes (PAD). The different bond pad voltage domains are indicated. All digital blocks have been omitted for clarity.}
\label{fig_Structures}
%\vspace{-1mm}
\end{figure}

To enable ${T_{chan}}$ characterization through a range of ${T_{amb}}$, gate thermometry is employed, in which the calibrated temperature dependence of the gate resistance ($R_{G}$, see Section~\ref{sect_Calibration}) is used as a temperature sensor~\cite{Triantopoulos19,Pavlidis16}. Kelvin connections to both top ($V_{GT}$, $I_{GT}$) and bottom ($V_{GB}$, $I_{GB}$) side of the H2 gate are therefore available to mitigate the impact of temperature dependence of back-end metals and parasitics on the resistance measurement (see Fig.~\ref{fig_Structures}).


The assumption was made that the gate and the channel are tightly thermally coupled, since only a thin (< \SI{3}{\nano \meter}) insulating layer separates them: $T_{G} = T_{chan}$.
\subsection{Diode Test Structures}
\label{sect_DiodeTestStructures}%
For the measurement of the thermal profile around the heaters, the substrate temperature is sensed by measuring the calibrated thermal dependency of the voltage drop ($V_A$) across $\mathrm{P^+}$/NWELL silicon diodes operated at a constant current $I_0$. A graphical representation of this structure can be seen in Fig.~\ref{fig_Structures}, comprising pad-accessible diodes and a multiplexed diode array.\\
\textbf{Pad-accessible diodes}: two diodes (D1 and D2 in Fig.~\ref{fig_Structures}) are placed in close vicinity to the heaters (one on each side), with connections directly available via bond pads, to measure

\begin{figure}[t!]
\setlength\fwidth{0.35\textwidth}
\centering
\hspace{1mm}
\includegraphics[angle=0]{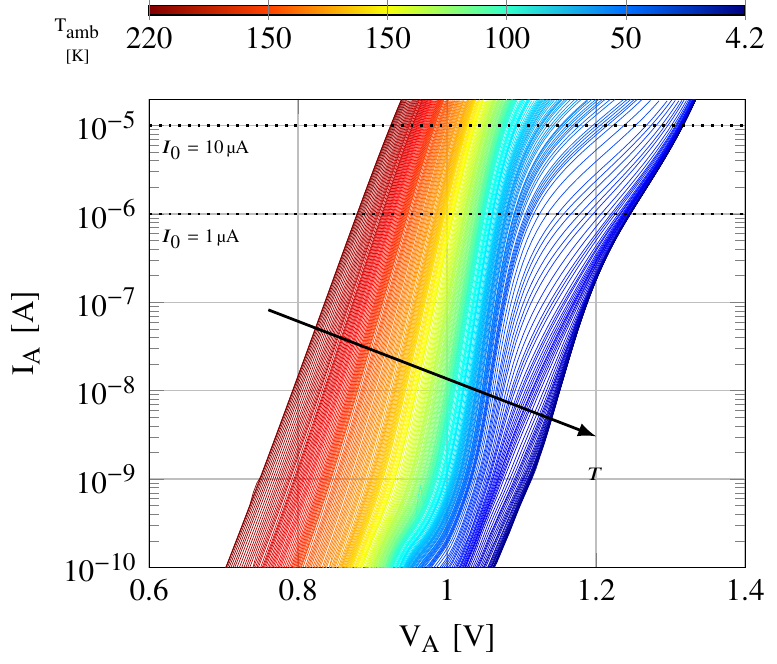}
\vspace{-1mm}
\caption{$I_A$-$V_A$ curves of pad-accessible diode D1 at ambient temperatures ($T_{amb}$) ranging from \SI{220}{\kelvin} down to \SI{4.2}{\kelvin}. The two horizontal cuts along $I_A=I_0$, from which the calibration curves are extracted, are indicated.}
\label{fig_IAVA}
%\vspace{-4.5mm}
\end{figure}%

the substrate temperature at small distances from the heaters with high spatial resolution (\SI{300}{\nano \meter}). Different combinations of heaters (H1/H2/H3) and diodes (D1/D2) allow for a total set of 6 distances: $d$ = \{1, 1.3, 1.6, 2, 2.3, 2.6\} \SI{}{\micro \meter}.\\
Because of the direct connection to the pads, these diodes have been used as benchmark to verify the correct operation of the pass gates in the multiplexed diode array.\\
\textbf{Multiplexed diode array}: a multiplexed array comprising 50 diodes to characterize the substrate temperature over larger distances, up to \SI{30}{\micro \meter} from the heaters, enables automatic characterization. Thick-oxide pass gates were employed to allow the diode potential to rise above the nominal supply voltage (\SI{1.1}{\volt}), required as $V_A|_{I_0}$ increases with decreasing temperature.\\
An array was placed on both sides of the heaters. The `dense' array (Fig.~\ref{fig_Structures} top right) comprises diodes placed at the minimum allowed distance, resulting in a spatial resolution of \SI{0.8}{\micro \meter} and is used for the actual measurements. The `sparse' array (Fig.~\ref{fig_Structures} top left) is a copy of the `dense' array with every other device removed, resulting in less contact/metal density compared to the latter array. By comparing the results from the two arrays, it can be verified if the metal/contact density significantly impacts the thermal profile due to heat-leakage via the biasing metal lines.

\subsection{Measurement Setup}
A photographic overview of the measurement setup can be seen in Fig.~\ref{fig_Setup}. The dies were glued and wire-bonded to ceramic DIP packages, which were fitted in a socket on a PCB mounted at the end of a dipstick (Fig.~\ref{fig_Setup}c). The PCB contains relays, enabling different configurations to be switched in and out during characterization. A Cernox type Resistance Temperature Detector (RTD) clamped to the package was used to measure $T_{amb}$.\\

\begin{figure}[t!]
\hspace{-6mm}%
\setlength\fwidth{0.35\textwidth}
\centering
\hspace{1mm}%
\includegraphics[angle=0]{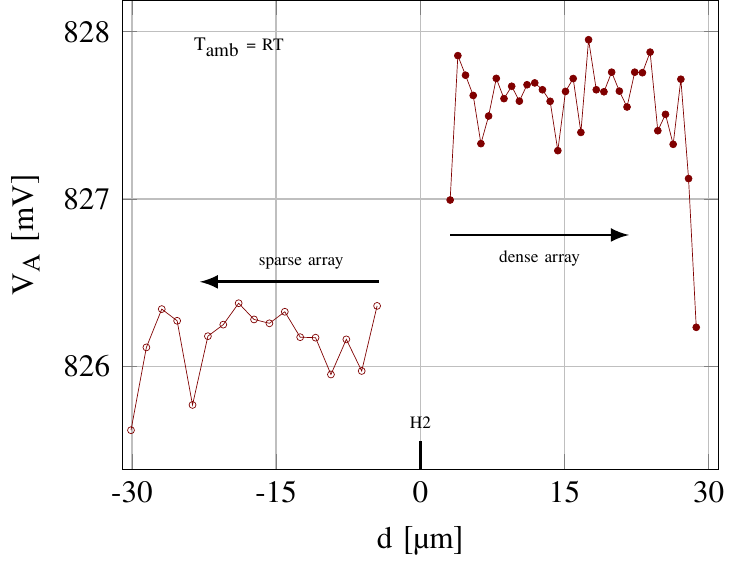}
\vspace{-1mm}%
\caption{Voltage drop ($V_A$) measured at $I_0 = \SI{10}{\micro \ampere}$ for diodes in the `dense' (filled dots) and `sparse' (open dots) array operated at RT with $P_H$ = 0 as a function of distance from center heater ($d$).}%
\label{fig_Variability}%
%\vspace{-4.5mm}
\end{figure}%
\begin{figure}[t!]
\setlength\fwidth{0.35\textwidth}
\centering
\hspace{1mm}
\includegraphics[angle=0]{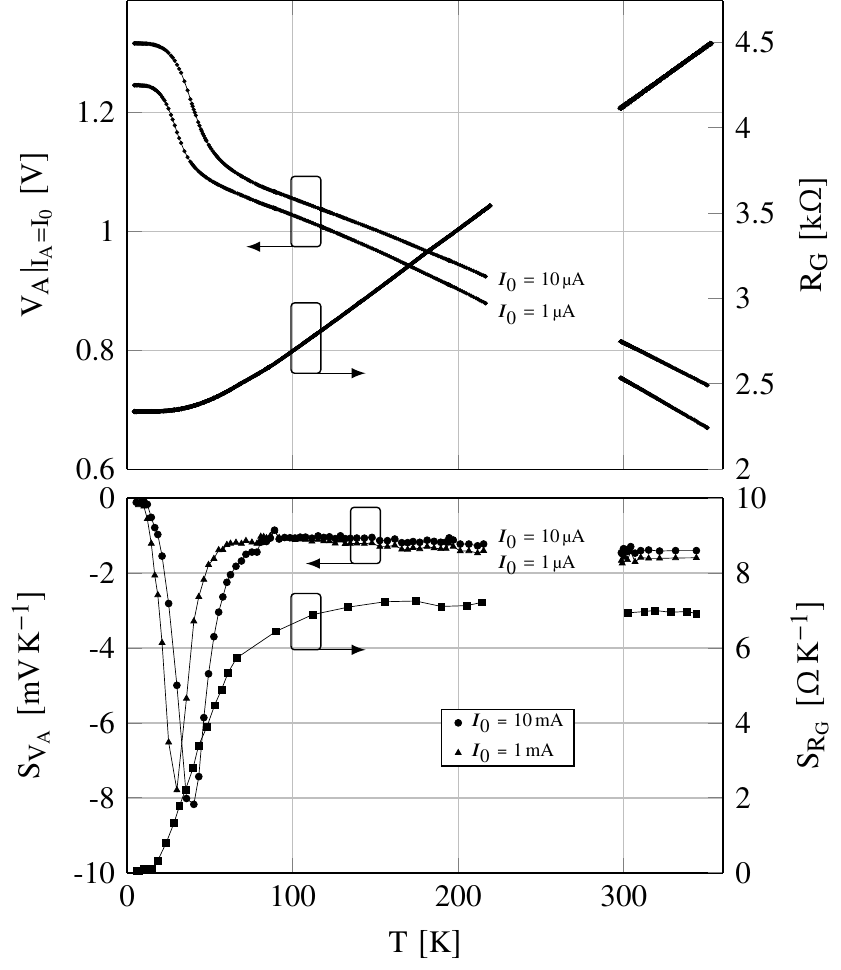}
\vspace{-1mm}
\caption{Calibration details. Top) calibration curves for the gate resistance ($R_{G}$) and the voltage drop across one of the pad-accessible diodes ($V_A$), the latter extracted at two different $I_0$ values, see also Fig.~\ref{fig_IAVA}. Bottom) temperature sensitivity of the gate resistance ($S_{R_G}$) and diode voltage ($S_{V_A}$) as a function of temperature ($T$) extracted from the calibration curves.}
\label{fig_CalibrationCurves}
%\vspace{-4.5mm}
\end{figure}%

Measurements at $T_{amb}$ $\geq$ RT were carried out by inserting the end of the dipstick into a V{\"o}tsch VTM7004 climate chamber (Fig.~\ref{fig_Setup}b). The PCB inside the climate chamber was enclosed by a metal box to improve thermal stability, thus reducing temperature drift/gradients over time (Fig.~\ref{fig_Setup}d).\\
The dipstick was inserted into a dewar containing LHe for the cryogenic measurements (Fig.~\ref{fig_Setup}a). The height of the sample above the LHe level modulates $T_{amb}$.\\
Electrical characterization was carried out by 3 Keithley 2636B SMUs.

\subsection{Calibration}
\label{sect_Calibration}
The temperature characteristics of both the gate resistor and the silicon diodes need to be calibrated before they can serve as temperature sensors. During calibration, the parameters of interest ($R_{G}$ and $V_A|_{I_0}$)  as a function of $T_{amb}$ are characterized while $T_{amb}$ is slowly varied with all heaters disabled.\\
For $T_{amb}$ $\geq$ RT, the climate chamber is used to generate a slowly varying $T_{amb}$: after warming up to $\approx \SI{350}{\kelvin}$, the climate chamber is switched off and allowed to (slowly) cool down while calibration takes place. Cryogenic calibration was carried out by manually lowering the dipstick into the dewar, cooling down the sample.\\
\textbf{Diode Calibration}: for the 2 pad-accessible diodes, the $I_A$-$V_A$ curves are measured as a function of $T_{amb}$. Deviation from ideal exponential behavior at cryogenic temperatures can be observed in Fig.~\ref{fig_IAVA}. From these curves, $V_A$ as a function of $T_{amb}$ is subsequently extracted by a horizontal cut along the line $I_A$ = $I_0$ as indicated in the figure. As recording the full $I_A$-$V_A$ characteristics for all 50 diodes in the array would take a prohibitive amount of time, $V_A$ is directly measured by forcing $I_A$ = $I_0$ for these devices. Since there is some variability present among different diodes, all diodes need to be calibrated individually, see Fig.~\ref{fig_Variability}.\\
An example of a diode calibration curve and the resulting temperature sensitivity can be observed in Fig.~\ref{fig_CalibrationCurves} top and bottom, respectively.\\
The de-facto standard value (for commercial diode temperature sensors) of $I_0$ = \SI{10}{\micro \ampere} was used for $T_{amb}$ $\geq$ RT. To increase the sensitivity at deep-cryogenic temperatures, the current bias was reduced to $I_0$ = \SI{1}{\micro \ampere} for $T_{amb}$ < \SI{300}{\kelvin}. The maximum temperature drift ($T_{drift}$) during a single $I_A$-$V_A$ characterization was \SI{0.6}{\kelvin} ($T_{amb}$ $\geq$ RT) and \SI{1}{\kelvin} ($T_{amb}$ < RT).\\
\textbf{Gate Calibration}: $R_{G}$ is measured by setting $V_{GB} = \SI{0}{\volt}$ and simultaneously  sweeping $V_{GT}$ from 0 to \SI{50}{\milli \volt}, see Fig.~\ref{fig_Structures}, while recording the current through the gate ($I_G$). Note that $V_{D}$ was left open to avoid any current and consequent heating in the device. $R_{G}$ is extracted from the slope of a first-order fit of the $I_G$-$V_{GT}$ characteristic. The full gate calibration curve can be observed in Fig.~\ref{fig_CalibrationCurves} top. The maximum $T_{drift}$ during a single $R_G$ characterization was \SI{0.25}{\kelvin} ($T_{amb}$ $\geq$ RT) and \SI{0.4}{\kelvin} ($T_{amb}$ < RT).\\
The gap between \SI{220}{\kelvin} and \SI{300}{\kelvin} results from limitations in the minimum and maximum attainable temperature of the climate chamber and LHe dewar, respectively.

\section{Experimental Results}
\label{sect_MeasurementsResultsAndAnalysis}
This section presents the measurement results, focusing on the experimental methods and discussing data validity. In-depth analysis and discussion of the reported data is given in Section~\ref{sect_Discussion}.
\subsection{Gate-resistance Measurements}
\label{sect_GateMeasurementAndAnalysis}
In the first step of the $T_{chan}$ characterization, the sample is brought to the target $T_{amb}$ by placing it at a certain height above the LHe level, or for RT measurements, by keeping it inside the (switched-off) climate chamber. When $T_{amb}$ stays within $\pm$ \SI{0.2}{\kelvin} of the set point, thermalization is assumed and different power levels are dissipated in the center heater ($P_{H2}$) by stepping $V_D$ in a staircase pattern: $V_D$ = \{0, 0.05, 0.1, ..., 1.1\} \SI{}{\volt}, while $V_{GB}$ and $V_{GT}$ are both set to \SI{1.1}{\volt}. Following each step in $V_D$, a \SI{10}{\second} delay was added to allow the structure to reach thermal equilibrium. $V_{GT}$ is subsequently swept from 1.1 to \SI{1.15}{\volt}, while both $I_D$ and $I_G$ are recorded. Finally, the routine

\begin{figure}[t!]
\setlength\fwidth{0.35\textwidth}
\centering
\hspace{1mm}
\includegraphics[angle=0]{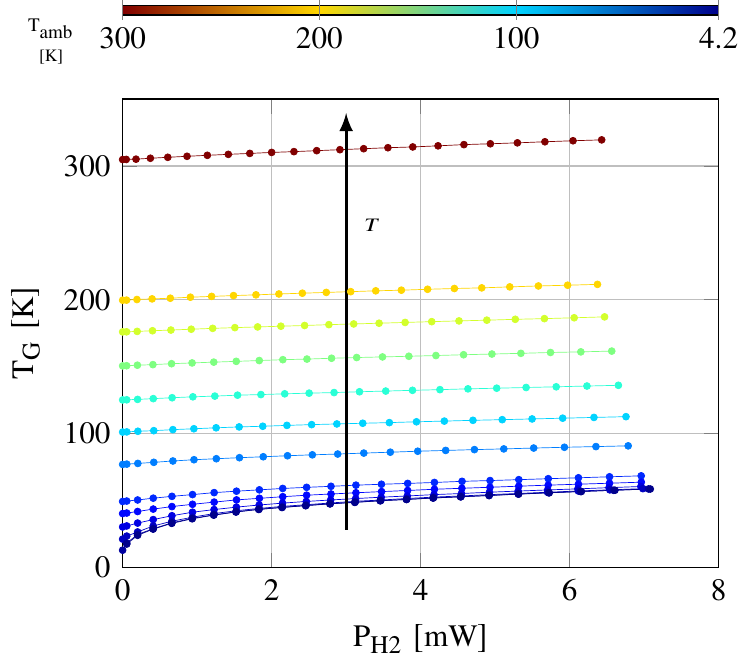}
\vspace{-1mm}
\caption{Absolute channel temperature ($T_G$) as a function of dissipated heater power ($P_{H2}$) at different ambient temperatures: $T_{amb}$ = \{4.2, 10, 20, 30, 40, 50, 75, 100, 125, 150, 175, 200, 300\} \SI{}{\kelvin}.}
\label{fig_AbsTgOverT}
%\vspace{-4.5mm}
\end{figure}%
\begin{figure}[t!]
\setlength\fwidth{0.35\textwidth}
\centering
\hspace{1mm}
\includegraphics[angle=0]{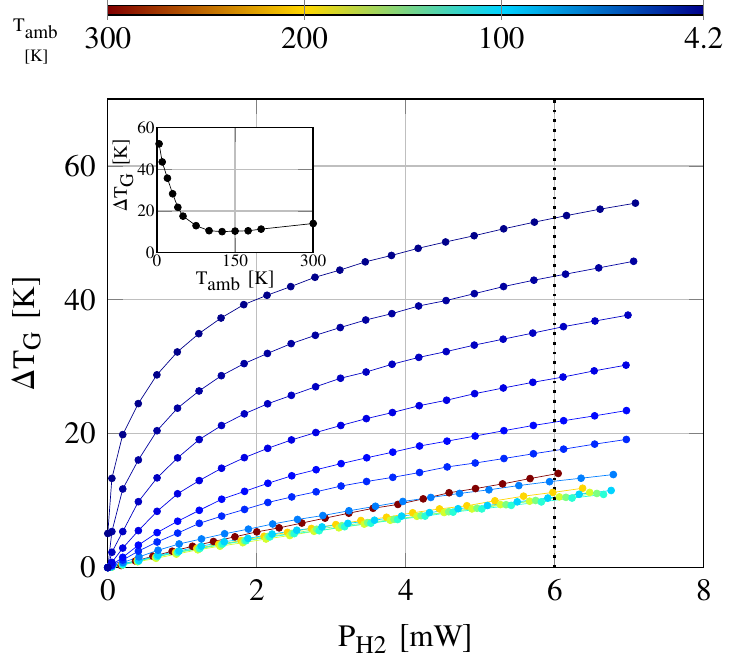}
\vspace{-1mm}
\caption{Channel self-heating ($\Delta T_G$) as a function of dissipated heater power ($P_{H2}$) for different ambient temperatures ($T_{amb}$). Inset indicates interpolated SH at fixed heater power ($P_{H2}$ = \SI{6}{\milli \watt}, see dashed line) as a function of $T_{amb}$. Plot derived from data in Fig.~\ref{fig_AbsTgOverT}.}
\label{fig_DeltaTgOverT}
%\vspace{-4.5mm}
\end{figure}%

described above is repeated for multiple $T_{amb}$.\\
$R_{G}$ is extracted from the $I_G$-$V_{GT}$ data as per the calibration routine described in Section~\ref{sect_Calibration}. The $T_{chan}$ is inferred from these extracted $R_{G}$ values by local Taylor-expansion of the calibration curve (Fig.~\ref{fig_CalibrationCurves} top) around that operating point.\\
As one side of the gate ($V_{GT}$) experiences a voltage change of \SI{50}{\milli \volt} during measurement, $I_D$ increases slightly. The dissipated power is therefore calculated using the mean $I_D$, $P_{H2} = V_D \cdot \bar{I}_D$, resulting in a maximum error of \SI{2.5}{\percent} in power.\\
The absolute $T_{chan}$ as a function of $P_{H2}$ for different $T_{amb}$ is plotted in Fig.~\ref{fig_AbsTgOverT}. The channel self-heating, $\Delta T_{chan}$, is derived from these data by subtracting the extracted temperature at $P_{H2} = 0$ for each $T_{amb}$ from the corresponding curve, as shown in Fig.~\ref{fig_DeltaTgOverT}. As the temperature sensitivity of $R_{G}$ drops to very low values for $T_{chan} < 11\SI{}{\kelvin}$, the RTD temperature reading was used for compensation instead of $T_{chan}|_{P_{H2}=0}$ for these curves.\\
In order to protect against sudden temperature changes in the helium vapour (caused by varying pressure in the building's helium recovery system), the $T_{amb}$ readings of the RTD are monitored: measurements are discarded when $T_{drift}$ exceeds $\pm$ \SI{0.5}{\kelvin} during a gate measurement at a single $V_D$ set point. In addition, these readings are also cross-checked with the extracted $T_{chan}|_{P_{H2}=0}$ as an additional safeguard.

\subsection{Diode Measurements}
\label{sect_DiodeMeasurementandAnalysis}
\begin{figure}[t!]
\setlength\fwidth{0.35\textwidth}
\centering
\hspace{-4mm}
\includegraphics[angle=0]{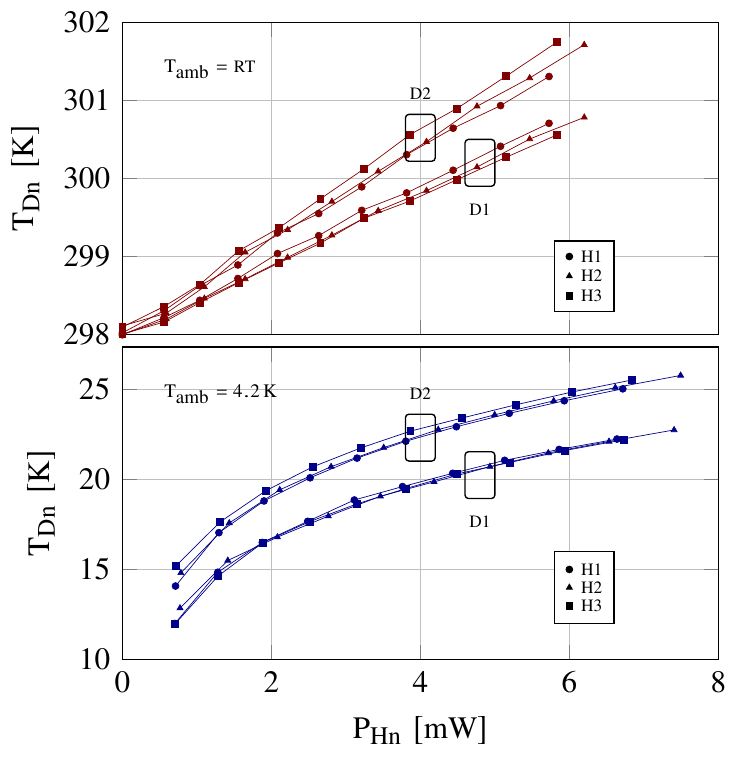}
\caption{Absolute diode temperature ($T_{Dn}$) measured with all 6 heater/diode combinations as a function of heater power ($P_{Hn}$) at both $T_{amb}$ = RT (top) and \SI{4.2}{\kelvin} (bottom). At \SI{4.2}{\kelvin} for $P_{Hn}$ below \SI{700}{\micro \watt}, readings are discarded due to limited temperature sensitivity, as explained in Section~\ref{subsect_Diode-BasedTemperatureSensing}.}
\label{fig_IndividualDiodeAbsToverT}
\end{figure}

\begin{figure}[t!]
\setlength\fwidth{0.35\textwidth}
\centering
%\hspace{1mm}
\includegraphics[angle=0]{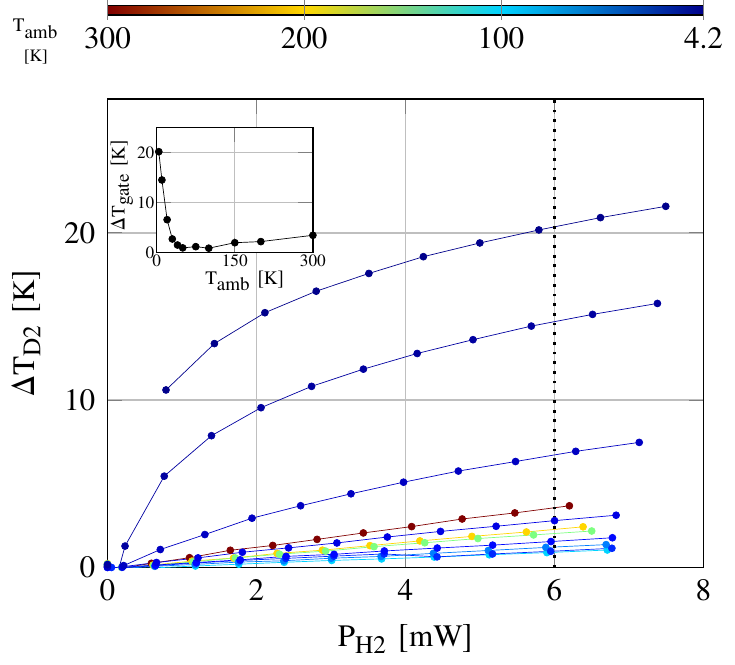}
\vspace{-1mm}
\caption{Diode temperature rise ($\Delta T_{D2}$) of D2 as a function of heater power ($P_{H2}$) at different ambient temperatures: $T_{amb}$ = \{4.2, 10, 20, 30, 40, 50, 75, 100, 150, 200, 300\} \SI{}{\kelvin}. Inset indicates interpolated temperature rise at fixed heater power ($P_{H2}$ = \SI{6}{\milli \watt}, see dashed line) as a function of $T_{amb}$.}
\label{fig_IndividualDiodeDToverT}
%\vspace{-4.5mm}
\end{figure}

The diode characterization is very similar to that of the gate (Section~\ref{sect_GateMeasurementAndAnalysis}); however, apart from H2, in this case, H1 and H3 can additionally be used as heaters.\\
The characterization and analysis of the diode measurements can again be split into two groups:\\
\textbf{Pad-accessible diodes}: $T_{amb}$, power dissipation and thermalization are handled as per the gate measurements. For each $V_D$ set point, $V_A$ is swept while $I_A$ and $I_D$ are recorded. These data are collected for all 6 combinations of H1, H2 or H3 with D1 or D2 over all $T_{amb}$ targets. From the $I_A$-$V_A$ data, $V_A|_{I_A = I_0}$ is extracted in the same manner as during diode calibration and the diode temperature ($T_{D}$) is inferred with the use of the individual diode calibration curves (Fig.~\ref{fig_CalibrationCurves} top).\\
The absolute $T_{D}$ for $T_{amb}$ = RT and \SI{4.2}{\kelvin} as a function of the enabled heater and the heater power can be seen in Fig.~\ref{fig_IndividualDiodeAbsToverT}.\\
Finally, the substrate heating (Fig.~\ref{fig_IndividualDiodeDToverT}) is calculated by compensating each absolute temperature curve in Fig.~\ref{fig_IndividualDiodeAbsToverT} with the corresponding temperature extracted at $P_{Hn}=0$, identical to the procedure followed in the channel SH analysis, while for $T_{amb} < \SI{10}{\kelvin}$ the temperature reading of the RTD was used.\\
\textbf{Multiplexed diode array}: the measurement and analysis of the diode array follow the same routine as the pad-accessible diodes described above; however, due to the large number of devices (50), resulting in an increased measurement time, some adaptions were implemented to mitigate long-term $T_{drift}$: since sub-\SI{}{\micro \meter} resolution is not required, only H2 was enabled; $V_A$ of each diode is directly measured by forcing $I_A$~=~$I_0$ and the number of $V_D$ set points was reduced to the set $V_D$ = \{0, 0.1, 0.2, ..., 1.1\} \SI{}{\volt}. With these measures in place, characterization of the full array still consumes a considerable amount of time, therefore $T_{drift}$ needs to be taken into account.\\


\begin{figure}[t!]
\setlength\fwidth{0.35\textwidth}
\centering
\hspace{1mm}
\includegraphics[angle=0]{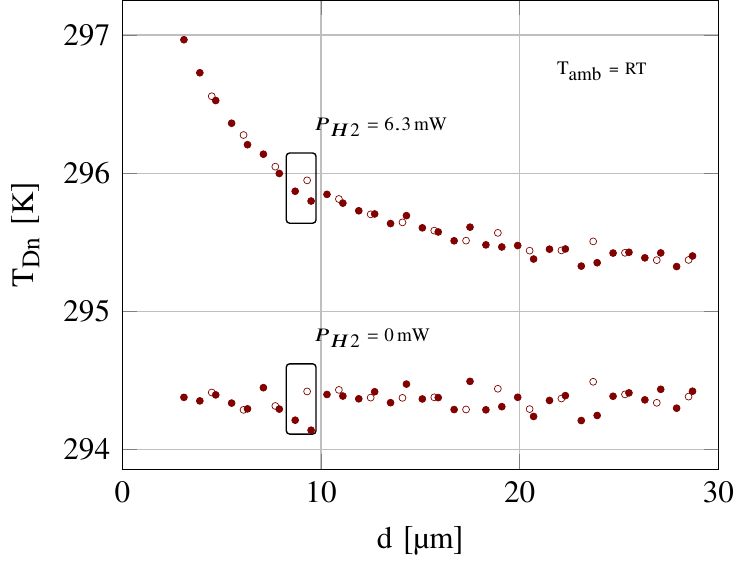}
\vspace{-1mm}
\caption{Substrate temperature measured by the diode arrays ($T_{Dn}$) as a function of distance from the center heater ($d$) at low ($P_{H2} = 0$) and high ($P_{H2} = \SI{6.3}{\milli \watt}$) heater power at $T_{amb}$ = RT. Data from the `dense' (filled dots) and `sparse' (open dots) arrays have been overlaid.}
\label{fig_ArrayAbsToverD_300K}
%\vspace{-4.5mm}
\end{figure}

Long-term temperature drift is only present for samples in helium vapour, caused by time-varying pressure in the building's helium recovery system.\\
To further minimize $T_{drift}$ impact, each diode in an array is fully characterized over all $V_D$ set points (power levels), before switching to adjacent devices. During characterization of a single diode, $T_{drift}$ is assumed to be small (comparable to that of the pad-accessible diodes) as the characterization time is relatively short: $\approx \SI{115}{\second}$. However, there is still a long-term $T_{drift}$ present between diode measurements, since a full array characterization takes $\approx \SI{97}{\minute}$. Therefore, the same compensation employed in the channel and pad-accessible diode characterization is applied here, which in this case, additionally auto-zeros the drift component between individual diode measurements.\\
The long-term drift is assumed to be small enough to maintain $T_{amb}$, however, large enough to distort the SH measurement, the effect of which is dependent on the height above the LHe. During the full array characterization, the RTD readings are therefore used to guard against too large short- and long-term $T_{drift}$. The allowed short-term drift (during single-diode measurements) is as per the pad-accessible diode characterization, while the long-term drift must stay within $\pm \SI{0.5}{\kelvin}$ of the target $T_{amb}$ for the data not to be discarded.\\
Extracted absolute substrate temperatures as a function of distance for $P_{H2}$ = 0 and $P_{H2} = \SI{6.3}{\milli \watt}$ measured at $T_{amb}$ = RT are plotted in Fig.~\ref{fig_ArrayAbsToverD_300K}.\\
Substrate heating as a function of distance at $P_{H2} = \SI{6}{\milli \watt}$ measured at different $T_{amb}$ can be observed in Fig.~\ref{fig_ArrayDeltaToverD}. The corresponding measured temperatures of the pad-accessible diodes and the channel have been added to the figure.\\
The temperature profiles associated with different heater powers at a fixed $T_{amb}$ = \SI{160}{\kelvin} are plotted in Fig.~\ref{fig_ArrayDTvsD_power_160K}, exemplifying the effect of $T_{drift}$ on a single diode measurement. 

\section{Self-Heating: Discussion, Modeling and Take-Aways}
\label{sect_Discussion}
\subsection{Diode-Based Temperature Sensing}
\label{subsect_Diode-BasedTemperatureSensing}
Deviation from exponential behavior in cryogenically operated diodes shown in Fig.~\ref{fig_IAVA}, are compatible with previous observations in literature~\cite{Shwarts08, Matthus19}. The sharp $V_A$ increase for $T_{amb}$ < \SI{50}{\kelvin} can be attributed to carrier freeze-out~\cite{Matthus19, Shwarts99, Ward03}, also present in diodes specifically designed for cryogenic temperature sensing~\cite{Courts15}.\\
Consistently and significantly lower $V_A|_{I_0=const}$ were found for diodes in the `sparse' array compared to the `dense' array as indicated in Fig.~\ref{fig_Variability}. Most likely these differences can be ascribed to a combination of two effects: 1) diodes in the `dense' array lie in a single continuous NWELL, while each `sparse' diode sits in its own well. This causes differences in doping densities, and subsequent electrical characteristics, through the Well Proximity Effect (WPE)~\cite{Hook05}. 2) the different Shallow Trench Isolation (STI) widths between diodes in the two arrays cause different mechanical stress to be present, altering the carrier transport parameters through the piezo-junction effect~\cite{Creemer01}.\\

\begin{figure}[t!]
\setlength\fwidth{0.35\textwidth}
\centering
\hspace{1mm}
\includegraphics[angle=0]{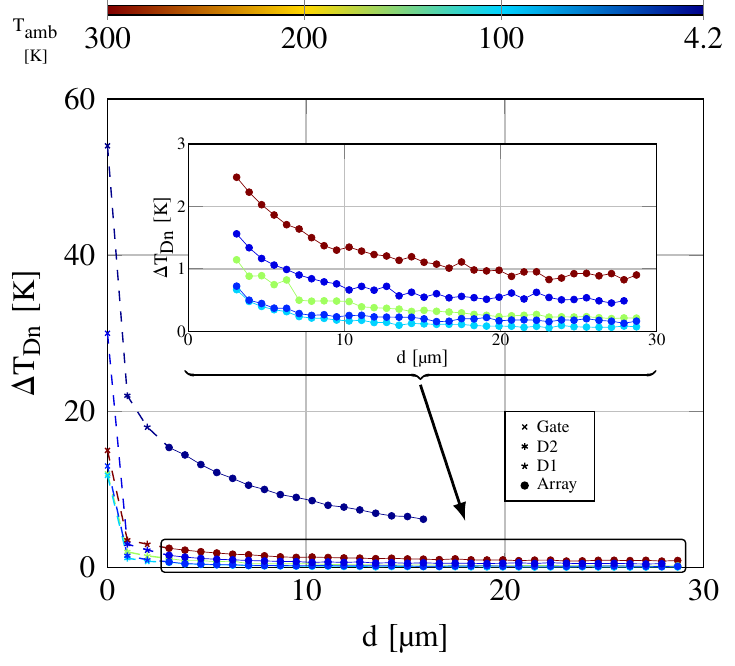}
\vspace{-1mm}
\caption{Substrate heating ($\Delta T_{Dn}$) measured by the `dense' array as a function of distance from the center heater ($d$) at high heater power ($P_{H2} = \SI{6.3}{\milli \watt}$) and different ambient temperatures: $T_{amb}$ = \{4.2, 30, 50, 100, 160, 300\} \SI{}{\kelvin}. Data from the gate and pad-accessible diode structures are also plotted. The inset indicates a zoomed-in plot of the diode array measurements with the \SI{4.2}{\kelvin} curve omitted for increased visibility.}
\label{fig_ArrayDeltaToverD}
%\vspace{-4.5mm}
\end{figure}%
\begin{figure}[t!]
\setlength\fwidth{0.35\textwidth}
\centering
\hspace{1mm}
\includegraphics[angle=0]{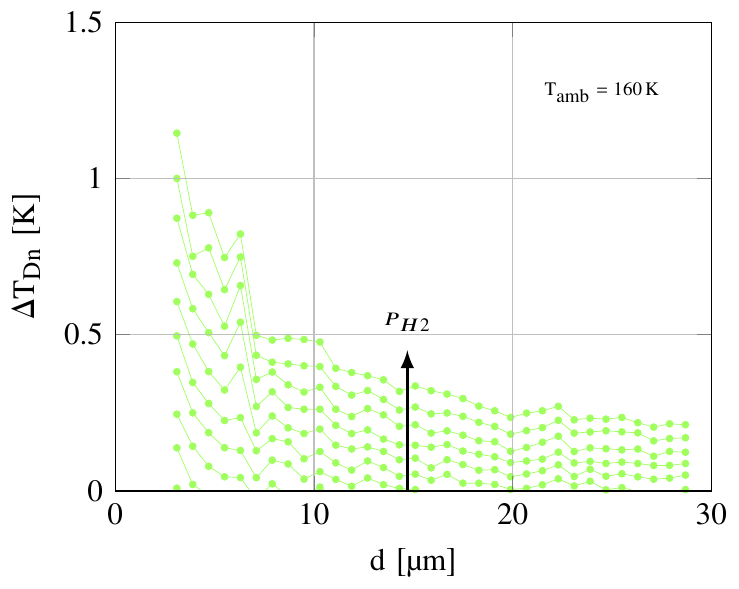} 
\vspace{-1mm}
\caption{Substrate heating measured by the `dense' array ($\Delta T_{Dn}$) as a function of distance to the center heater ($d$) at different heater power levels: $P_{H2} = \{0.6, 1.2, 1.7, 2.3, 3.0, 3.6, 4.3, 5.0, 5.8, 6.6\}$ \SI{}{\milli \watt}. $T_{amb}$ = \SI{160}{\kelvin}. Only curves corresponding to $P_{H2}$ values causing significant heating are plotted. $T_{amb}$ was selected for clearly showing the effects of $T_{drift}$, similar curves were found at other $T_{amb}$.}
\label{fig_ArrayDTvsD_power_160K}
\vspace{-4.5mm}
\end{figure}%

The calibration curves of $V_A$ and $R_{G}$ in Fig.~\ref{fig_CalibrationCurves} show a near-constant temperature sensitivity of \SI{-1.2}{\milli \volt / \kelvin} and \SI{0.18}{\percent / \kelvin} for $T_{amb}$ > \SI{50}{\kelvin}, respectively. Freeze-out causes a large increase in $V_A$ sensitivity below this temperature, however, sensitivity drops to a very low value for $T_{amb}$ < \SI{10}{\kelvin}, compatible with measurements in~\cite{Ward03}. $I_0$ was reduced to \SI{1}{\micro \ampere} during cryogenic measurements to mitigate the latter effect~\cite{Shwarts08}, the value being a trade-off between improved sensitivity and increased impact of array leakage at deep-cryogenic temperatures~\cite{Shwarts00}. The temperature sensitivity of $R_G$ decreases to a value close to zero for $T_{amb}$ < \SI{11}{\kelvin}, in-line with metal-like behavior~\cite{Alt13}. For $T_{amb}$ > \SI{50}{\kelvin}, carrier transport is limited by phonon scattering, exhibiting a positive temperature coefficient (PTC), while below this temperature transport becomes increasingly limited by impurity scattering, which being temperature independent, prevents further resistance decrease.\\
Due to the diminishing temperature sensitivity of both the diodes and gate resistance, the reading of $T_{D}$ and $T_{G}$ have been excluded for temperatures below \SI{10}{\kelvin} and \SI{11}{\kelvin}, respectively. However, due to the rapid temperature increase to values above \SI{10}{\kelvin} already at low $P_H$ for $T_{amb}$ = \SI{4.2}{\kelvin}, this results in the loss of only a small part of the data (see Fig.~\ref{fig_AbsTgOverT} and \ref{fig_IndividualDiodeAbsToverT}).\\ 
SH due to biasing of the temperature sensors is insignificant for the range of interest ($P_H$ > \SI{1}{\milli \watt}), as it has been simulated in COMSOL to be below \SI{8}{\micro \watt} for the diodes, resulting in $\Delta T_{max}$ < \SI{20}{\milli \kelvin}; and below \SI{1}{\micro \watt} for the gate resistor, resulting in $\Delta T_{max}$ < \SI{10}{\milli \kelvin}, respectively.

\subsection{Channel Temperature Sensing}
From the $T_{chan}$ measurements in Fig.~\ref{fig_AbsTgOverT}, an agreement between the RTD temperature reading and extracted $T_{chan}|_{P_{H2}=0}$ was found, indicating the correct operation of the setup through the full temperature range. Larger SH at equal $P_{H2}$ can be observed for lower temperatures in Fig.~\ref{fig_DeltaTgOverT}: $\Delta T_{chan} \approx$ \SI{14}{\kelvin} (RT) vs $\Delta T_{chan} \approx$ \SI{52}{\kelvin} (\SI{4.2}{\kelvin}) at $P_{H2}$ = \SI{6}{\milli \watt}. For $T_{amb}$ < \SI{100}{\kelvin}, $\Delta T_{chan}$ is highly non-linear with respect to dissipated power, resulting in large SH for low $P_{H2}$ in this temperature range, also observed by~\cite{Gutierrez92} and~\cite{Artanov21}.
As shown in the inset of Fig.~\ref{fig_DeltaTgOverT}, the SH behavior from RT down to \SI{4.2}{\kelvin} for a given $P_{H2}$ shows a decrease down to \SI{125}{\kelvin}, below which its effect starts to increase again, exhibiting a dramatic increase below $T_{amb}$ = \SI{75}{\kelvin}. 
This behavior hints to a temperature-dependent $R_{th}$, with a minimum at $\approx$ \SI{100}{\kelvin}, also shown in other works~\cite{Gutierrez92, Hidalga00}. Previously reported values of the minimum lie between \SI{77}{\kelvin} and \SI{250}{\kelvin} and have been attributed to the impact of parasitic $R_{th}$ (package, glue, etc.) dominating at these low temperatures in bulk CMOS. The crowding of $T_{chan}$ between \SI{40}{\kelvin} and \SI{60}{\kelvin} for deep-cryogenic temperatures, visible in Fig.~\ref{fig_AbsTgOverT} for heating power above \SI{1}{\milli \watt}, is a direct consequence of this $R_{th}$ behavior: below the $R_{th}$ minimum, $R_{th}$ has a negative temperature coefficient (NTC), which impedes SH more and more as $T_{chan}$ approaches the minimum. The implications of this observed effect are further discussed in Section~\ref{sect_SHImpactonCryo-CMOS}.\\
Comparing the SH magnitude extracted in this work with literature, much higher SH was found in SOI technology at comparable power densities~\cite{Triantopoulos19}. As the main $R_{th}$ in such technology is determined by SiO$_2$, which exceeds that of Si by $\approx$ 2 orders of magnitude, a large difference in SH is expected. Regarding bulk technology, in which no BOX exists, the geometry and area of MOSFET devices significantly impact SH. Far lower SH was observed in a large square heater in bulk technology~\cite{Hidalga00}, which has significantly more enclosing area and hence a much lower $R_{th}$ to the surrounding silicon compared to the wide/short devices measured in this work. Also, the power density is orders of magnitude less compared to that in this work. The values published on a bulk device with an aspect ratio better resembling the structures characterized in this work, but with much larger $W$ and $L$, show a slightly smaller SH effect. The structure in question had $\approx 100 \times$ larger area~\cite{Gutierrez93} and considerably lower power densities. The preliminary work done by~\cite{Artanov21} shows values that compare very well with the measurements presented here, although no geometrical details are given.\\ 
These results stress the importance of geometry on SH, which is why in this work a transistor geometry comparable to the ones employed in practical cryo-CMOS circuits was chosen.

\subsection{Spatial Thermal Measurements}
Observations in line with the previous two sections can be made for the pad-accessible diodes. Both diodes show a smaller $\Delta T_{D}$ compared to the $\Delta T_{chan}$ at identical conditions, as the effect of heating falls off rapidly with a $1/d^n$-law (with $d$ the distance to the heater and $n$ a factor between 1 and 2): at $T_{amb}$ = \SI{4.2}{\kelvin}, \SI{30}{\kelvin} less $\Delta T_{D}$ was measured \SI{1}{\micro \meter} from the heater compared to $\Delta T_{chan}$ itself, see Fig.~\ref{fig_IndividualDiodeAbsToverT} and~\ref{fig_IndividualDiodeDToverT}. Again, larger $\Delta T_{D}$ at cryogenic temperatures compared to RT was observed: $\Delta T_{D} \approx \SI{3.5}{\kelvin}$ (RT) vs $\Delta T_{diode} \approx \SI{21}{\kelvin}$ (\SI{4.2}{\kelvin}) at $P_H$ = \SI{6}{\milli \watt}, measured at \SI{1}{\micro \meter} from the heater.\\
All 6 diode/heater combinations are distinguishable at both RT and \SI{4.2}{\kelvin} in Fig.~\ref{fig_IndividualDiodeAbsToverT}, with $T_{D2}$ > $T_{D1}$, as D2 is closer to the heater than D1. The $\Delta T_D$ as a function of the enabled heater is flipped between D1 and D2, reflecting the mirror symmetry of the structure (see Fig.~\ref{fig_Structures} top).\\
In Fig.~\ref{fig_IndividualDiodeDToverT}, a similar behavior as in the $T_{chan}$ measurements can be observed in the pad-accessible diodes below \SI{30}{\kelvin}. The inset clearly shows the same behavior: a decreasing $\Delta T_D$ with decreasing $T_{amb}$ with a minimum at $\approx \SI{100}{\kelvin}$, which compares well with the channel measurement.\\ 
Additional cryogenic effects were observed in the heaters, see Fig.~\ref{fig_IndividualDiodeAbsToverT}. At $T_{amb}$ = \SI{4.2}{\kelvin}, the power in H1-H3 increases by \SI{17}{\percent} to \SI{20}{\percent} at equal bias conditions compared to RT, attributed to the improved mobility, resulting in an increased $I_D$ and $P_H$. At equal $T_{amb}$, H2 was able to dissipate consistently more power compared to H1 and H3. H2 is effectively shielded from STI stress by adjacent devices (H1 and H3), which alters carrier transport parameters (and thus $I_D$ and $P_H$) through the piezo-resistive effect~\cite{Creemer01}.\\
Another interesting observation on the SH structure is the heat propagation from the heaters to the `dense' and 'sparse' diode arrays. As seen in Fig.~\ref{fig_ArrayAbsToverD_300K}, there is good agreement between the $T_D$ in both arrays, indicating no significant effects of metal/STI density on the thermal transport for these measurements, as was described in Section~\ref{sect_DiodeTestStructures}. The readings at $P_{H2} = 0$ correlate well with the RTD readings and an agreement within $\pm \SI{0.25}{\kelvin}$ between both `dense' and `sparse' diodes was found at both low and high $P_{H2}$; the latter indicates a stable $T_{amb}$ and a successful calibration. The temperature mismatch between the two arrays is mainly due to the large time span between individual measurements, as each array is fully characterized before switching to the other. Compatibility of the array data with both channel and pad-accessible diode measurements can be seen, the shape corresponding to simulations shown by~\cite{Hidalga00}.\\
The substrate $\Delta T$ falls off with the distance $d$ from the heater for all measured $T_{amb}$, following a similar shape to $\Delta T$ measured at RT, see Fig.~\ref{fig_ArrayDeltaToverD}. At $T_{amb}$ = \SI{4.2}{\kelvin}, the observable $d$ range is limited as the substrate temperature drops below \SI{10}{\kelvin} for $d > \SI{15}{\micro \meter}$, as discussed previously. The $\Delta T_{D}$ evolution over $T_{amb}$ matches the pad-accessible diode data, e.g. a minimum at $\approx$ \SI{100}{\kelvin}. At $T_{amb}$ = \SI{4.2}{\kelvin} severe substrate heating was observed, as much as \SI{7}{\kelvin}, measured \SI{15}{\micro \meter} from the heater dissipating $P_{H2}$ = \SI{6.5}{\milli \watt}.\\
Substrate heating at $T_{amb}$ = \SI{160}{\kelvin} as a function of $P_{H2}$ (Fig.~\ref{fig_ArrayDTvsD_power_160K}) uncovers detectable substrate heating \SI{30}{\micro \meter} from the heater at $P_{H2}$ > \SI{3.6}{\milli \watt}, while negligible heating is observed at $P_{H2} \leq$ \SI{0.6}{\milli \watt}. A \SI{0.1}{\kelvin} short-term $T_{drift}$ is visible, impacting the diode measurement at $d = \SI{6.3}{\micro \meter}$.

\subsection{Ultra-Wide-Temperature Self-Heating Model}
\begin{figure}[t!]
\setlength\fwidth{0.35\textwidth}
\centering
\hspace{1mm}
\includegraphics[angle=0]{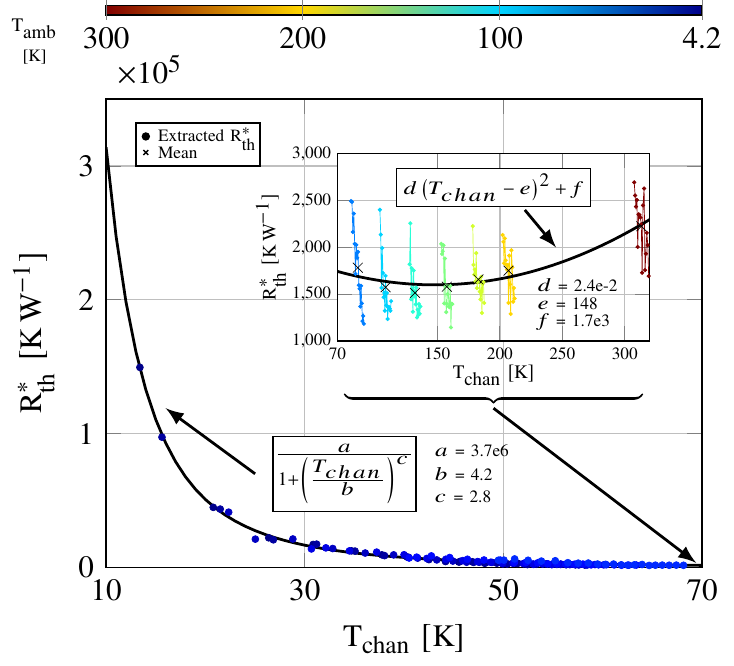} 
\vspace{-1mm}
\caption{Extracted differential thermal resistance ($R^*_{th}$) as a function of channel temperature ($T_{chan}$) derived from data shown in Fig.~\ref{fig_AbsTgOverT}. Data from different ambient temperatures ($T_{amb}$) are indicated by colors. The inset shows the extracted data at $T_{chan}$ > \SI{75}{\kelvin} for visibility. Fitting parameter values are indicated in the figure.}
\label{fig_RthOverT}
%\vspace{-4.5mm}
\end{figure}

\begin{figure}[!t]
\subfloat{\setlength\fwidth{1\textwidth}
	\hspace{-2mm}%
     	\setlength\fwidth{0.125\textwidth}%
	\includegraphics[angle=0]{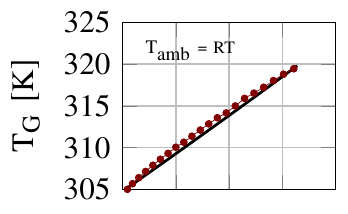}
	\hspace{-2mm}%
	\setlength\fwidth{0.125\textwidth}%
	\includegraphics[angle=0]{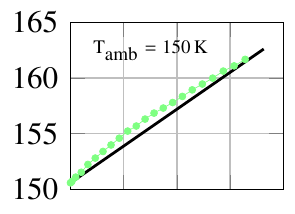}
	\hspace{-2mm}%
	\setlength\fwidth{0.125\textwidth}%
	\includegraphics[angle=0]{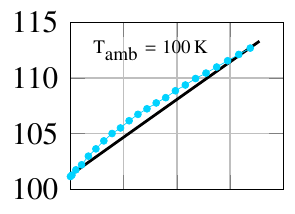}}\\[-1mm] 
\subfloat{\setlength\fwidth{1\textwidth}
	\hspace{-2mm}%
     	\setlength\fwidth{0.1245\textwidth}%
	\includegraphics[angle=0]{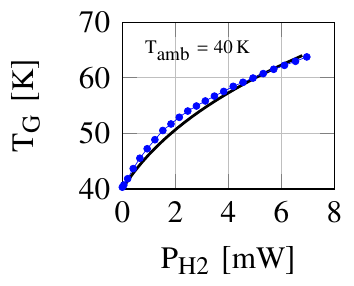}
	\hspace{-1.15mm}%
	\setlength\fwidth{0.125\textwidth}%
	\includegraphics[angle=0]{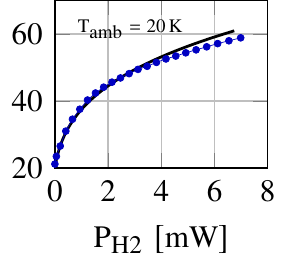}
	\hspace{-1.3mm}%
	\setlength\fwidth{0.125\textwidth}%
	\includegraphics[angle=0]{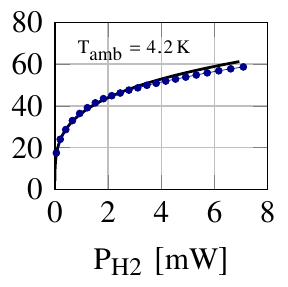}}
	\vspace{-2mm}%
	\caption{Absolute gate temperature ($T_G$) as a function of heater power ($P_{H2}$) at different ambient temperatures ($T_{amb}$). Model prediction (solid lines) vs measured data (dotted lines). Note: y-axes not equally scaled.}
	\label{fig_Model}
   \end{figure}

In order to make the IC design work-flow cryo-SH aware, SH was modeled via a similar approach as in~\cite{Triantopoulos19}, but for bulk CMOS. First, the differential thermal resistance ($R^*_{th} = d \Delta T_{chan}/dP_H$) has been calculated from data in Fig.~\ref{fig_DeltaTgOverT}, and plotted as a function of absolute channel temperature ($T_{chan} = T_{amb} + \Delta T_{chan}$) in Fig.~\ref{fig_RthOverT}. The extracted $R^*_{th}$ at $T_{amb}$ < \SI{50}{\kelvin} partially overlap, proving the validity of the measured channel SH. Since SH in bulk is far less pronounced compared to SOI, and even less at higher temperatures, the $T_{chan}$ range for $T_{amb}$ > \SI{50}{\kelvin} is limited and gaps appear in the $R^*_{th}$ curve. The previously discussed minimum and rapid increase in SH at deep-cryogenic temperatures are also reflected in this curve. The shape of the $R^*_{th}$ curve, and in particular the deviation from the expected $R^*_{th}$ valley around \SI{40}{\kelvin}, is compatible with the one shown in~\cite{Hidalga00}. This curve illustrates that a single function is able to describe the complete $R^*_{th}$ behavior of this structure over the full temperature range from RT down to \SI{4.2}{\kelvin}, and can thus be employed to model SH over $P_H$ and $T_{amb}$. As the data at deep-cryogenic temperatures are similarly shaped to those in~\cite{Triantopoulos19}, but deviates at higher temperatures (containing a minimum, not monotonically decreasing), $R^*_{th}$ was split into two regions, only to aid fitting. For $T_{chan} \leq$ \SI{70}{\kelvin} Eq.~(\ref{eqn_Rth})~\cite{Triantopoulos19} was used:

\begin{equation}
R^*_{th} = \frac{R^*_{th0}}{1 + \left( \frac{T}{T_0} \right)^n},
\label{eqn_Rth}
\end{equation}
\\
while for $T_{chan}$ > \SI{70}{\kelvin} a simple parabolic function was fitted to the data to capture the minimum and the PTC behavior, both shown in Fig.~\ref{fig_RthOverT}. Finally, these two fitted functions and Eq.~(\ref{eqn_model})~\cite{Triantopoulos19} were used to predict SH as a function of $T_{amb}$ and $P_H$.

\begin{equation}
P = \int_{0}^{\Delta T} \frac{d\Delta T'}{R^*_{th}(T_{amb} + \Delta T')}
\label{eqn_model}
\end{equation}
\\
The resulting models for various $T_{amb}$ are plotted in Fig.~\ref{fig_Model}. These plots show that the very simple Eq.~(\ref{eqn_model}) is capable of successfully predicting SH over the full $T_{amb}$ range from \SI{4.2}{\kelvin} up to RT, including both the linear and square-root-like behavior, with < \SI{3}{\kelvin} error in the 0 to \SI{7}{\milli \watt} $P_H$ range.\\

\subsection{SH Impact on Cryo-CMOS Circuits}
\label{sect_SHImpactonCryo-CMOS}
While the cryo-SH data and modeling presented above could enable the next steps in reliable cryo-CMOS design, conclusions on the impact of circuit behavior can already be drawn. Although SH is indeed severe for $T_{amb}$ below \SI{50}{\kelvin} (see Fig.~\ref{fig_DeltaTgOverT}), $T_{chan}$ will not exceed an absolute temperature above \SI{60}{\kelvin} even for a dissipated 

\begin{figure}[t!]
\setlength\fwidth{0.35\textwidth}
\centering
\hspace{1mm}
\includegraphics[angle=0]{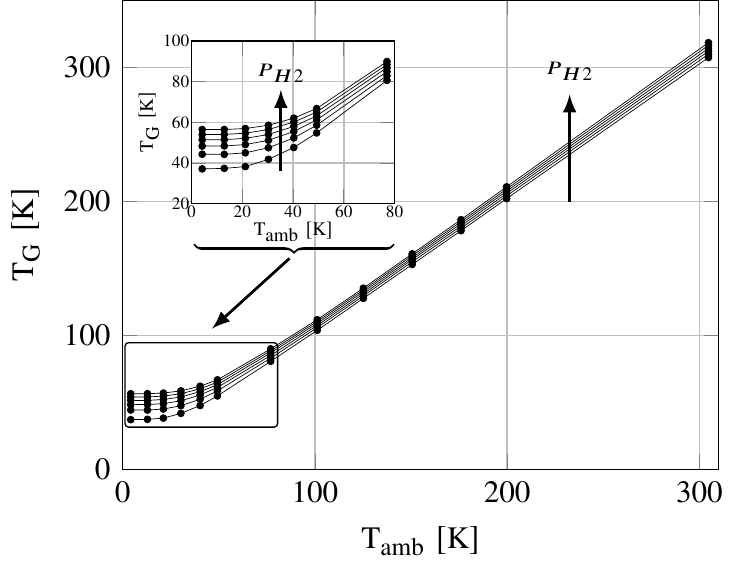}
\vspace{-1mm}
\caption{Gate temperature ($T_G$) as a function of ambient temperature ($T_{amb}$) for different H2 power levels: $P_{H2}$ = \{1, 2, ..., 6\} \SI{}{\milli \watt}. These data are interpolations  based on  data shown in Fig.~\ref{fig_AbsTgOverT}. A zoomed-in plot of the saturation region is shown in the inset.}
\label{fig_TchanVsTamb}
%\vspace{-4.5mm}
\end{figure}

power of a few \SI{}{\milli \watt}'s, due to the minimum in the thermal resistance, as clearly highlighted by replotting the data from Fig.~\ref{fig_AbsTgOverT} in Fig.~\ref{fig_TchanVsTamb}. Since the key transistor parameters, such as threshold voltage, current factor and subthreshold slope, as well as passive-device characteristics, such as $R_G$ (see Fig.~\ref{fig_CalibrationCurves}), were shown to saturate at temperatures below \SI{50}{\kelvin}~\cite{Hart20_1,Hart20_2}, SH would not significantly impact the circuit bias conditions and its dynamic performance, causing a relative temperature insensitivity in this regime. However, the increase in circuit temperature by tens of degree Kelvin above $T_{amb}$ can significantly degrade the noise performance for a thermal-noise-limited circuit, although it is still unclear whether temperature-independent shot-noise may be the main limitation in transistor’s noise performance at deep-cryogenic temperatures~\cite{Chen21}. Moreover, since the exact position of the thermal-resistance minimum cannot be fully attributed to the thermal properties of silicon, but heavily depends on the die thermalization, such as the package, the $T_{chan}$ may vary due different positions on the die or boundary conditions of the die with the surrounding enclosure. Devices at tens of \SI{}{\micro \meter} distance from each other can still experience significant thermal cross-talk at deep-cryogenic temperatures even at moderate power levels ($P$ > \SI{4}{\milli \watt}) within at least a radius of \SI{30}{\micro \meter} (see Fig.~\ref{fig_ArrayDeltaToverD} and~\ref{fig_ArrayDTvsD_power_160K}). This directly translates into layout guidelines to properly space power-hungry devices from noise-sensitive circuits and precision circuits for which matching is a major consideration.
 
\section{Conclusion}
\label{sect_Conclusion}
A 40-nm CMOS test structure was fabricated and characterized for a comprehensive evaluation of self-heating in bulk CMOS technology in the ambient temperature range from \SI{300}{\kelvin} down to \SI{4.2}{\kelvin}. The temperature rise was measured both in the MOSFET channel through the change in the gate resistance, and in the surrounding silicon substrate by a linear array of diodes operating as sensors.\\
Severe self-heating was observed at deep-cryogenic ambient temperatures, resulting in a channel temperature rise exceeding \SI{40}{\kelvin} for a dissipated power of only \SI{2}{\milli \watt} at a \SI{4.2}{\kelvin} ambient temperature. Although the thermal conductivity of silicon is relatively low at very low temperatures, the absolute channel temperature does not exceed \SI{60}{\kelvin} even for significantly higher power, due to the thermal resistance for a typical MOSFET, which has minimum above \SI{70}{\kelvin}. This effect was confirmed by extracting the device thermal resistance from measured data at different temperatures and modeling it with a simple analytical expression able to predict channel temperatures over the full ambient temperature range from deep-cryogenic to room temperature.\\
The spatial propagation of SH results in a rise in substrate temperature detectable and quantifiable at a distance of \SI{30}{\micro \meter} from the heater.\\ 
The thorough characterization of nanometer bulk-CMOS devices at cryogenic temperatures is of paramount importance for the design of the integrated control electronics for quantum processors. For achieving first-time-right silicon, it is imperative to simulate the circuit at the actual operating temperature rather than assume the ambient temperature. Towards that goal, the results and modeling presented in this work will contribute towards the full self-heating-aware IC design-flow required for the reliable design and operation of cryo-CMOS circuits.

\section*{Acknowledgment}
The authors would like to thank Dr. H.P.~Tuinhout for the helpful discussions and Intel for funding.

\bibliographystyle{IEEEtran}
\bibliography{bibliography}

\begin{thebibliography}{10}
\providecommand{\url}[1]{#1}
\csname url@samestyle\endcsname
\providecommand{\newblock}{\relax}
\providecommand{\bibinfo}[2]{#2}
\providecommand{\BIBentrySTDinterwordspacing}{\spaceskip=0pt\relax}
\providecommand{\BIBentryALTinterwordstretchfactor}{4}
\providecommand{\BIBentryALTinterwordspacing}{\spaceskip=\fontdimen2\font plus
\BIBentryALTinterwordstretchfactor\fontdimen3\font minus
  \fontdimen4\font\relax}
\providecommand{\BIBforeignlanguage}[2]{{%
\expandafter\ifx\csname l@#1\endcsname\relax
\typeout{** WARNING: IEEEtran.bst: No hyphenation pattern has been}%
\typeout{** loaded for the language `#1'. Using the pattern for}%
\typeout{** the default language instead.}%
\else
\language=\csname l@#1\endcsname
\fi
#2}}
\providecommand{\BIBdecl}{\relax}
\BIBdecl

\bibitem{Vandersypen17}
L.~M.~K. Vandersypen, H.~Bluhm, J.~S. Clarke, A.~S. Dzurak, R.~Ishihara,
  A.~Morello, D.~J. Reilly, L.~R. Schreiber, and M.~Veldhorst, ``{I}nterfacing
  {S}pin {Q}ubits in {Q}uantum {D}ots and {D}onors—{H}ot, {D}ense, and
  {C}oherent,'' \emph{NPJ Quantum Information}, vol.~3, pp. 34--44, 08 2017.

\bibitem{Vandijk19}
J.~{van Dijk}, E.~Charbon, and F.~Sebastiano, ``{T}he {E}lectronic {I}nterface
  for {Q}uantum {P}rocessors,'' \emph{Microprocessors and Microsystems},
  vol.~66, pp. 90--101, 2019.

\bibitem{Google19}
F.~Arute, K.~Arya, R.~Babbush, D.~Bacon, J.~Bardin, R.~Barends, R.~Biswas,
  S.~Boixo, F.~Brandao, D.~Buell, B.~Burkett, Y.~Chen, Z.~Chen, B.~Chiaro,
  R.~Collins, W.~Courtney, A.~Dunsworth, E.~Farhi, B.~Foxen, and J.~Martinis,
  ``{Q}uantum {S}upremacy {U}sing a {P}rogrammable {S}uperconducting
  {P}rocessor,'' \emph{Nature}, vol. 574, pp. 505--510, 10 2019.

\bibitem{Sebastiano17}
F.~Sebastiano, H.~Homulle, B.~Patra, R.~Incandela, J.~van Dijk, L.~Song,
  M.~Babaie, A.~Vladimirescu, and E.~Charbon, ``{C}ryo-{CMOS} {E}lectronic
  {C}ontrol for {S}calable {Q}uantum {C}omputing,'' in \emph{2017 54th
  ACM/EDAC/IEEE DAC}, 2017, pp. 1--6.

\bibitem{Incandela18}
R.~M. Incandela, L.~Song, H.~Homulle, E.~Charbon, A.~Vladimirescu, and
  F.~Sebastiano, ``{C}haracterization and {C}ompact {M}odeling of {N}anometer
  {CMOS} {T}ransistors at {D}eep-{C}ryogenic {T}emperatures,'' \emph{IEEE
  JEDS}.

\bibitem{Ekanayake10}
S.~R. {Ekanayake}, T.~{Lehmann}, A.~S. {Dzurak}, R.~G. {Clark}, and
  A.~{Brawley}, ``{C}haracterization of {SOS-CMOS} {FET}s at {L}ow
  {T}emperatures for the {D}esign of {I}ntegrated {C}ircuits for {Q}uantum
  {B}it {C}ontrol and {R}eadout,'' \emph{IEEE TED}, vol.~57, no.~2, pp.
  539--547, Feb 2010.

\bibitem{Beckers18}
A.~Beckers, F.~Jazaeri, and C.~Enz, ``{C}haracterization and {M}odeling of
  28-nm {B}ulk {CMOS} {T}echnology {D}own to 4.2 {K},'' \emph{IEEE JEDS},
  vol.~6, pp. 1007--1018, 2018.

\bibitem{Hart20_1}
P.~A. ’t Hart, M.~Babaie, E.~Charbon, A.~Vladimirescu, and F.~Sebastiano,
  ``{C}haracterization and {M}odeling of {M}ismatch in {C}ryo-{CMOS},''
  \emph{IEEE JEDS}, vol.~8, pp. 263--273, 2020.

\bibitem{Hart20_2}
P.~A. 't~Hart, M.~Babaie, E.~Charbon, A.~Vladimirescu, and F.~Sebastiano,
  ``{S}ubthreshold {M}ismatch in {N}anometer {CMOS} at {C}ryogenic
  {T}emperatures,'' \emph{IEEE JEDS}, vol.~8, pp. 797--806, 2020.

\bibitem{Patra20}
B.~Patra, M.~Mehrpoo, A.~Ruffino, F.~Sebastiano, E.~Charbon, and M.~Babaie,
  ``{C}haracterization and {A}nalysis of {O}n-{C}hip {M}icrowave {P}assive
  {C}omponents at {C}ryogenic {T}emperatures,'' \emph{IEEE JEDS}, vol.~8, pp.
  448--456, 2020.

\bibitem{Glassbrenner64}
C.~J. Glassbrenner and G.~A. Slack, \emph{Phys. Rev.}, vol. 134, pp.
  A1058--A1069, May 1964.

\bibitem{Vandijk20}
J.~P.~G. Van~Dijk, B.~Patra, S.~Subramanian, X.~Xue, N.~Samkharadze, A.~Corna,
  C.~Jeon, F.~Sheikh, E.~Juarez-Hernandez, B.~P. Esparza, H.~Rampurawala, B.~R.
  Carlton, S.~Ravikumar, C.~Nieva, S.~Kim, H.-J. Lee, A.~Sammak, G.~Scappucci,
  M.~Veldhorst, L.~M.~K. Vandersypen, E.~Charbon, S.~Pellerano, M.~Babaie, and
  F.~Sebastiano, ``{A} {S}calable {C}ryo-{CMOS} {C}ontroller for the {W}ideband
  {F}requency-{M}ultiplexed {C}ontrol of {S}pin {Q}ubits and {T}ransmons,''
  \emph{IEEE JSSC}, vol.~55, no.~11, pp. 2930--2946, 2020.

\bibitem{Gutierrez93}
E.~Gutierrez-D., L.~Deferm, and G.~Declerck, ``{E}xperimental {D}etermination
  of {S}elf-{H}eating in {S}ubmicrometer {MOS} {T}ransistors {O}perated in a
  {L}iquid-{H}elium {A}mbient,'' \emph{IEEE EDL}, vol.~14, no.~3, pp. 152--154,
  1993.

\bibitem{Asheghi98}
M.~Asheghi, M.~Touzelbaev, K.~Goodson, Y.~Leung, and S.~Wong,
  ``{T}emperature-{D}ependent {T}hermal {C}onductivity of {S}ingle-{C}rystal
  {S}ilicon {L}ayers in {SOI} {S}ubstrates,'' \emph{J HEAT TRANSFER}, vol. 120,
  02 1998.

\bibitem{Sesnic72}
S.~Sesnic and G.~Craig, ``{T}hermal {E}ffects in {JFET} and {MOSFET} {D}evices
  at {C}ryogenic {T}emperatures,'' \emph{IEEE TED}, vol.~19, no.~8, pp.
  933--942, 1972.

\bibitem{Foty87}
D.~Foty and S.~Titcomb, ``{T}hermal {E}ffects in n-{C}hannel {E}nhancement
  {MOSFET}'s {O}perated at {C}ryogenic {T}emperatures,'' \emph{IEEE TED},
  vol.~34, no.~1, pp. 107--113, 1987.

\bibitem{Foty89}
D.~Foty, ``{T}hermal {E}ffects in p-{C}hannel {MOSFET}s at {L}ow
  {T}emperatures,'' \emph{IEEE TED}, vol.~36, no.~8, pp. 1542--1544, 1989.

\bibitem{Artanov21}
A.~Artanov, A.~Cabrera-Galicia, A.~Kruth, C.~Degenhardt, E.~Gutierrez-D,
  D.~Durini, and S.~Waasen, ``{S}elf-{H}eating {E}ffect in 65nm {CMOS}
  {T}echnology (poster),'' in \emph{2021 WOLTE14}, 2021.

\bibitem{Gutierrez93_2}
E.~A. {Gutiérrez D.}, L.~Deferm, and G.~Declerck, ``{S}elfheating {E}ffects in
  {S}ilicon {R}esistors {O}perated at {C}ryogenic {A}mbient {T}emperatures,''
  \emph{SSE}, vol.~36, no.~1, pp. 41--52, 1993.

\bibitem{Hidalga00}
F.~De~la Hidalga, M.~Deen, and E.~Gutierrez, ``{T}heoretical and {E}xperimental
  {C}haracterization of {S}elf-{H}eating in {S}ilicon {I}ntegrated {D}evices
  {O}perating at {L}ow {T}emperatures,'' \emph{IEEE TED}, vol.~47, no.~5, pp.
  1098--1106, 2000.

\bibitem{Jomaah95}
J.~Jomaah, G.~Ghibaudo, and F.~Balestra, ``Analysis and {M}odeling of
  {S}elf-{H}eating {E}ffects in {T}hin-{F}ilm {SOI} {MOSFET}s as a {F}unction
  of {T}emperature,'' \emph{SSE}, vol.~38, no.~3, pp. 615--618, 1995.

\bibitem{Triantopoulos19}
K.~Triantopoulos, M.~Cassé, S.~Barraud, S.~Haendler, E.~Vincent, M.~Vinet,
  F.~Gaillard, and G.~Ghibaudo, ``{S}elf-{H}eating {E}ffect in {FDSOI}
  {T}ransistors {D}own to {C}ryogenic {O}peration at 4.2 {K},'' \emph{IEEE
  TED}, vol.~66, no.~8, pp. 3498--3505, 2019.

\bibitem{Hidalga97}
F.~De~la Hidalga-W. and E.~Guttierrez-D., ``{T}he n-{MOS} {T}ransistor as a
  {L}ow {T}emperature {T}hermometer,'' 1997.

\bibitem{Gutierrez92}
E.~A. Gutiérrez-D., L.~Deferm, S.~Decoutere, and G.~Declerck, ``{E}xperimental
  {D}etermination of {S}elfheating in {S}ilicon {R}esistors {O}perated at
  {C}ryogenic {T}emperatures,'' \emph{Microelectronic Engineering}, vol.~19,
  no.~1, pp. 865--868, 1992.

\bibitem{Gutierrez97}
E.~Gutierrez-D., J.~De~la Hidalga-W., M.~Deen, and S.~Koshevaya, ``{A}n
  {A}lternative {M}ethod to {M}onitor and {C}ontrol the {IC} {T}emperature in
  the 4.2-77 {K} {R}ange,'' in \emph{27th ESSDERC}, 1997, pp. 436--439.

\bibitem{Kiene21}
G.~Kiene, A.~Catania, R.~Overwater, P.~Bruschi, E.~Charbon, M.~Babaie, and
  F.~Sebastiano, ``{A} 1{GS}/s 6-to-8b 0.5m{W}/{Q}ubit {C}ryo-{CMOS} {SAR}
  {ADC} for {Q}uantum {C}omputing in 40nm {CMOS},'' in \emph{2021 IEEE ISSCC},
  vol.~64, 2021, pp. 214--216.

\bibitem{Prabowo21}
B.~Prabowo, G.~Zheng, M.~Mehrpoo, B.~Patra, P.~Harvey-Collard, J.~Dijkema,
  A.~Sammak, G.~Scappucci, E.~Charbon, F.~Sebastiano, L.~M.~K. Vandersypen, and
  M.~Babaie, ``{A} 6-to-8{GH}z 0.17m{W}/{Q}ubit {C}ryo-{CMOS} {R}eceiver for
  {M}ultiple {S}pin {Q}ubit {R}eadout in 40nm {CMOS} {T}echnology,'' in
  \emph{2021 IEEE ISSCC}, vol.~64, 2021, pp. 212--214.

\bibitem{Ruffino21}
A.~Ruffino, Y.~Peng, T.-Y. Yang, J.~Michniewicz, M.~F. Gonzalez-Zalba, and
  E.~Charbon, ``{A} {F}ully-{I}ntegrated 40-nm 5-6.5 {GH}z {C}ryo-{CMOS}
  {S}ystem-on-{C}hip with {I/Q} {R}eceiver and {F}requency {S}ynthesizer for
  {S}calable {M}ultiplexed {R}eadout of {Q}uantum {D}ots,'' in \emph{2021 IEEE
  ISSCC}, vol.~64, 2021, pp. 210--212.

\bibitem{Park21}
J.-S. Park, S.~Subramanian, L.~Lampert, T.~Mladenov, I.~Klotchkov, D.~J.
  Kurian, E.~Juarez-Hernandez, B.~Perez-Esparza, S.~R. Kale, K.~T. Asma~Beevi,
  S.~Premaratne, T.~Watson, S.~Suzuki, M.~Rahman, J.~B. Timbadiya, S.~Soni, and
  S.~Pellerano, ``{A} {F}ully {I}ntegrated {C}ryo-{CMOS} {S}o{C} for {Q}ubit
  {C}ontrol in {Q}uantum {C}omputers {C}apable of {S}tate {M}anipulation,
  {R}eadout and {H}igh-{S}peed {G}ate {P}ulsing of {S}pin {Q}ubits in {I}ntel
  22nm {FFL} {F}in{FET} {T}echnology,'' in \emph{2021 IEEE ISSCC}, vol.~64,
  2021, pp. 208--210.

\bibitem{Pavlidis16}
G.~Pavlidis, S.~Pavlidis, E.~R. Heller, E.~A. Moore, R.~Vetury, and S.~Graham,
  ``{C}haracterization of {A}l{G}a{N}/{G}a{N} {HEMT}s {U}sing {G}ate
  {R}esistance {T}hermometry,'' \emph{IEEE Transactions on Electron Devices},
  vol.~64, no.~1, pp. 78--83, 2017.

\bibitem{Shwarts08}
Y.~M. Shwarts, M.~M. Shwarts, and S.~V. Sapon, ``{A} {N}ew {G}eneration of
  {C}ryogenic {S}ilicon {D}iode {T}emperature {S}ensors,'' in \emph{2008
  ASDAM}, 2008, pp. 239--242.

\bibitem{Matthus19}
C.~D. Matthus, L.~Di~Benedetto, M.~Kocher, A.~J. Bauer, G.~D. Licciardo,
  A.~Rubino, and T.~Erlbacher, ``{F}easibility of 4{H}-{S}i{C} p-i-n {D}iode
  for {S}ensitive {T}emperature {M}easurements {B}etween 20.5 {K} and 802
  {K},'' \emph{IEEE JSEN}, vol.~19, no.~8, pp. 2871--2878, 2019.

\bibitem{Shwarts99}
Y.~Shwarts, V.~Borblik, N.~Kulish, V.~Sokolov, M.~Shwarts, and E.~Venger,
  ``{S}ilicon {D}iode {T}emperature {S}ensor {W}ithout a {K}ink of the
  {R}esponse {C}urve in {C}ryogenic {T}emperature {R}egion,'' \emph{Sensors and
  Actuators A: Physical}, vol.~76, no.~1, pp. 107--111, 1999.

\bibitem{Ward03}
R.~Ward, W.~Dawson, L.~Zhu, R.~Kirschman, O.~Mueller, M.~Hennessy, E.~Mueller,
  R.~Patterson, J.~Dickman, and A.~Hammoud, ``{P}ower {D}iodes for {C}ryogenic
  {O}peration,'' in \emph{IEEE 34th PESC}, vol.~4, 2003, pp. 1891--1896 vol.4.

\bibitem{Courts15}
S.~S. Courts, ``{A} {S}tandardized {D}iode {C}ryogenic {T}emperature {S}ensor
  for {A}erospace {A}pplications,'' \emph{Cryogenics}, vol.~74, pp. 172--179,
  2016, 2015 Space Cryogenics Workshop.

\bibitem{Hook05}
T.~Hook, J.~Brown, and X.~Tian, ``{P}roximity {E}ffects and {VLSI} {D}esign,''
  06 2005, pp. 167 -- 170.

\bibitem{Creemer01}
J.~Creemer, F.~Fruett, G.~Meijer, and P.~French, ``{T}he {P}iezojunction
  {E}ffect in {S}ilicon {S}ensors and {C}ircuits and its {R}elation to
  {P}iezoresistance,'' \emph{IEEE JSEN}, vol.~1, no.~2, pp. 98--, 2001.

\bibitem{Shwarts00}
Y.~Shwarts, V.~Sokolov, M.~Shwarts, I.~Fedorov, and E.~Venger, ``{A}dvanced
  {S}ilicon {D}iode {T}emperature {S}ensors with {M}inimized {S}elf-{H}eating
  and {N}oise for {C}ryogenic {A}pplications,'' in \emph{ASDAM}, 2000, pp.
  351--354.

\bibitem{Alt13}
A.~Alt and C.~Bolognesi, ``{T}emperature {D}ependence of {A}nnealed and
  {N}onannealed {HEMT} {O}hmic {C}ontacts {B}etween 5 and 350 {K},'' \emph{IEEE
  TED}, vol.~60, pp. 787--792, 02 2013.

\bibitem{Chen21}
X.~Chen, H.~Elgabra, C.-H. Chen, J.~Baugh, and L.~Wei, ``{E}stimation of
  {MOSFET} {C}hannel {N}oise and {N}oise {P}erformance of {CMOS} {LNA}s at
  {C}ryogenic {T}emperatures,'' in \emph{2021 IEEE ISCAS}, 2021, pp. 1--5.

\end{thebibliography}

\begin{IEEEbiography}[{\includegraphics[width=1in,height=1.25in,clip,keepaspectratio]{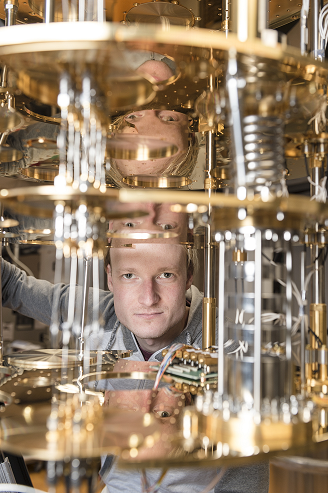}}]{P.A. `t Hart}
received the M.Sc. in Electrical Engineering from the Delft University of Technology, The Netherlands, in 2014. For his thesis he worked in the Electronic Components, Materials and Technology department (ECTM) and the Electronic Instrumentation Laboratory, where he did characterization of the piezojunction effect in bipolar transistors. In 2017 he joined the group of Edoardo Charbon where he is currently pursuing the Ph.D degree characterizing and modeling cryogenic MOSFETs. His research interests include: low temperature electronics, modeling and characterization.
\end{IEEEbiography}
\begin{IEEEbiography}[{\includegraphics[width=1in,height=1.25in,clip,keepaspectratio]{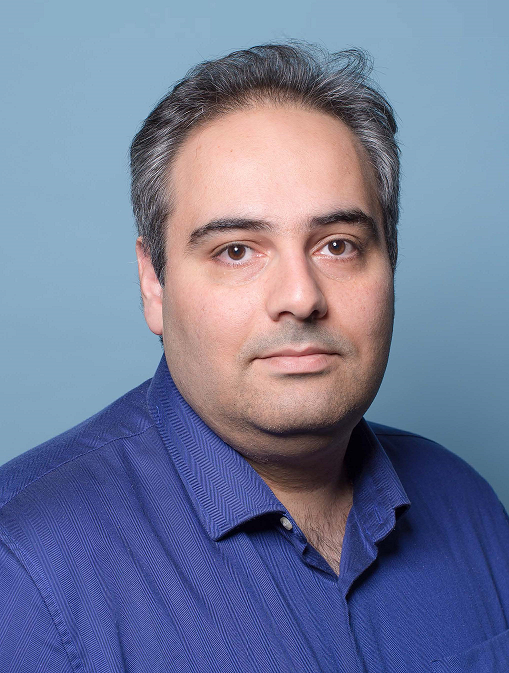}}]{M. Babae}(S'12--M'16)
received the Ph.D. degree (cum laude) in electrical engineering from the Delft University of Technology, Delft, The Netherlands, in 2016. 
In 2006, he joined the Kavoshcom Research and Development Group, Tehran, where he was involved in designing wireless communication systems. From 2009 to 2011, he was a CTO of that company. From 2014 to 2015, he was a Visiting Scholar Researcher with the Berkeley Wireless Research Center, Berkeley, CA, USA. In 2016, he joined the Delft University of Technology, where he is currently an Assistant Professor (tenured). His current research interests include RF/millimeter-wave integrated circuits and systems for wireless communications, and cryogenic electronics for quantum computation.
Dr. Babaie has been a committee member of the Student Research Preview (SRP) of the IEEE International Solid-State Circuits Conference (ISSCC) since 2017. He is currently serving on the technical program committee of the IEEE European Solid-State Circuits Conference (ESSCIRC). He was a co-recipient of the 2015–2016 IEEE Solid-State Circuits Society Pre-Doctoral Achievement Award, and the 2019 IEEE ISSCC Best Demo Award. In 2019, he received the Veni award from the Netherlands Organization for Scientific Research (NWO).
\end{IEEEbiography}
\begin{IEEEbiography}[{\includegraphics[width=1in,height=1.25in,clip,keepaspectratio]{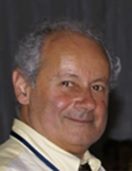}}]{A. Vladimirescu} (F'17) received the M.S. and Ph.D. degrees in EECS from the University of California, Berkeley, where he was a key contributor to the SPICE simulator, releasing the SPICE2G6 production-level SW in 1981. He pioneered electrical simulation on parallel computers with the CLASSIE simulator as part of his PhD. He is the author of "The SPICE Book" published by J. Wiley and Sons. 
For many years Andrei was R$\&$D director leading the design and implementation of innovative software and hardware Electronic Design Automation products for Analog Devices Inc., Daisy Systems, Analog Design Tools, Valid Logic and Cadence.
Currently he is Professor involved in research projects at the University of California at Berkeley, Delft University of Technology and the Institut Sup\'erieur d'\'Electronique de Paris, ISEP, as well as consultant to industry. His research activities are in the areas of ultra-low-voltage (ULV) CMOS, design, simulation and modeling of circuits with new devices and circuits for quantum computing.
Andrei is an IEEE Fellow.
\end{IEEEbiography}
\begin{IEEEbiography}[{\includegraphics[width=1in,height=1.25in,clip,keepaspectratio]{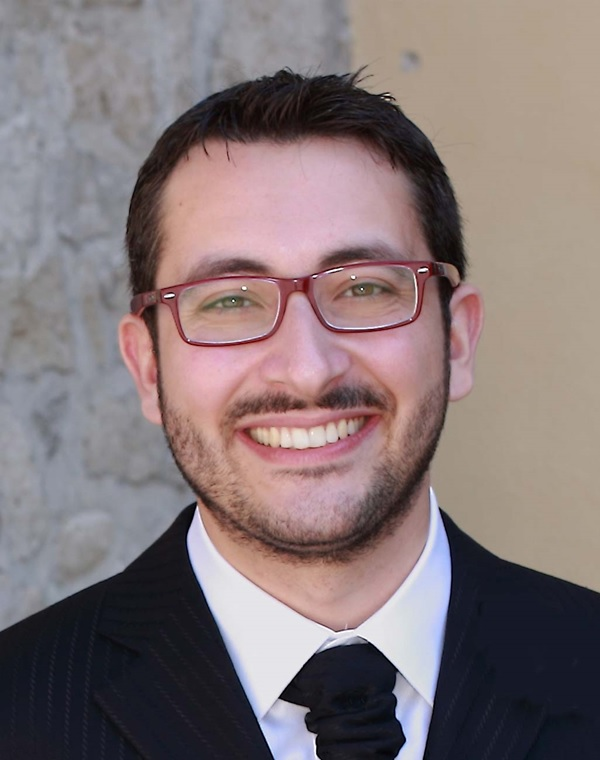}}]{F. Sebastiano}(S’09–M’10–SM’17) received the B.Sc. (cum laude) and M.Sc. (cum laude) degrees in electrical engineering from University of Pisa, Italy, in 2003 and 2005, respectively, the M.Sc. degree (cum laude) from Sant’Anna school of Advanced Studies, Pisa, Italy, in 2006 and the Ph.D. degree from Delft University of Technology, The Netherlands, in 2011.
From 2006 to 2013, he was with NXP Semiconductors Research in Eindhoven, The Netherlands, where he conducted research on fully integrated CMOS frequency references, deep-submicron temperature sensors and area-efficient interfaces for magnetic  sensors. In 2013, he joined Delft University of Technology, where he is currently an Associate Professor and the Research Lead of the Quantum Computing Division of QuTech. He has authored or co-authored one book, 11 patents and over 80 technical publications. His main research interests are cryogenic electronics, quantum computing, sensor read-outs and fully integrated frequency references.
Dr. Sebastiano is on the technical program committee of the ISSCC, the IEEE RFIC Symposium, and the IMS, and he is currently serving as an Associate Editor of the of IEEE Transactions on VLSI. He was co-recipient of the 2008 ISCAS Best Student Paper Award, the 2017 DATE best IP award, and the ISSCC 2020 Jan van Vessem Award for Outstanding European Paper. He has served as Distinguished Lecturer of the IEEE Solid-State Circuit Society. 

\end{IEEEbiography}

\end{document}